\documentclass{aa}
\usepackage{graphicx}
\usepackage{txfonts}
\usepackage{natbib}
\bibpunct{(}{)}{;}{a}{}{,} 

\begin{document}

   \title{Free Core Nutation observed by VLBI}


   \author{H. Kr\'asn\'a \inst{1}
          \and
          J. B\"ohm \inst{1}
          \and
          H. Schuh\inst{2}
          }

   \institute{Vienna University of Technology, Department of Geodesy and Geoinformation,
Research Group Advanced Geodesy, Gusshausstrasse 27-29, 1040 - Vienna, Austria\\
              \email{hana.krasna@tuwien.ac.at}
         \and
             Helmholtz-Zentrum Potsdam, DeutschesGeoForschungsZentrum GFZ, Department 1: Geodesy and Remote Sensing, Potsdam, Germany\\
             }

   \date{Received March 27, 2013; accepted April 24, 2013}

  \abstract
   {}
   {The signature of free core nutation (FCN) is found in the motion of the celestial intermediate pole in the celestial reference frame and in the resonance behaviour of the frequency-dependent Earth tidal displacement in its diurnal band. We focus on estimation of the FCN parameters, i.e. the period and amplitude.}
   {We run several global adjustments of 27~years of very long baseline interferometry (VLBI) data (1984.0 - 2011.0) to determine the FCN period from partial derivatives of the VLBI observables with respect to the FCN as contained in the nutation of the celestial intermediate pole and in the solid Earth tidal displacement in the diurnal band. Finally, we estimate the FCN period by a global adjustment from both phenomena simultaneously, which has not been done before.}
   {We find that our estimate of the FCN period of $-431.18~\pm~0.10$ sidereal days slightly deviates from the conventional value of $-431.39$ sidereal days. Additionally, we present our empirical model of the FCN with variable amplitude and phase compatible with the estimated period.}
   {}

   \keywords{methods: data analysis --
            techniques: interferometric --
                astrometry --
                reference systems --
                Earth
               }

   \maketitle
%

\section{Introduction}
\begin{figure}
\centering
\includegraphics[width=\hsize]{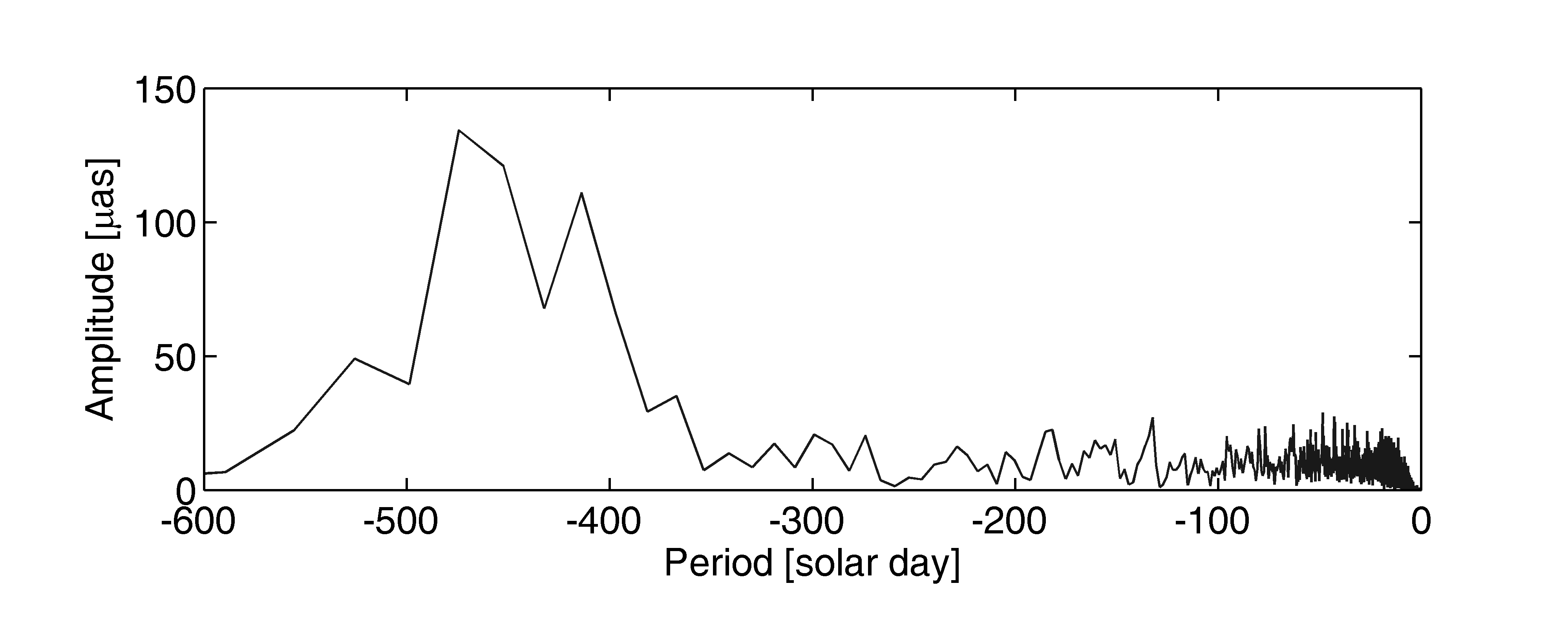}
  \caption{Fourier spectrum of CPO (dX + idY) estimated with software VieVS with respect to the IAU 2006/2000A precession-nutation model.}
     \label{Fig_dXdY_fft}
\end{figure}

The rotating Earth has several free rotational modes, one of them being free core nutation (FCN). This normal mode is caused by the fact that the ellipsoidal liquid core inside the visco-elastic Earth's mantle rotates around an axis which is slightly misaligned with the axis of the mantle. In the celestial reference frame (CRF) it is visible as a retrograde motion of the Earth figure axis with a period of about 431 days and has an amplitude of about 100 microarcseconds~\citep{Mathews02, Vondrak05, Lambert07a}. Since there are no models available which could predict this free motion with its time-varying excitation and damping, it is not included in the precession-nutation model of the Earth axis adopted in the current International Earth Rotation and Reference Systems Service (IERS) Conventions 2010~\citep{iers10}. Therefore, the dominant part of the residuals between the direction of the celestial intermediate pole (CIP) in the CRF as observed by very long baseline interferometry (VLBI) and the direction modelled by the very accurate precession-nutation model, adopted by the International Astronomical Union (IAU), IAU 2006/2000A \citep{Mathews02, Capitaine03} is caused by the FCN. In the terrestrial reference frame (TRF) the motion is observed at a period of about one day and is designated as nearly diurnal free wobble (NDFW). At this frequency, i.e. in the diurnal band, there is a strong resonance between the NDFW and the solid Earth tidal displacement. In this work we focus on estimation of the FCN period from the nutation motion of the Earth's axis in space and also from the resonance behaviour in the diurnal tidal band.\\
There have been several investigations on the FCN period from VLBI data in the past. For example, spectral and wavelet techniques have been applied to the celestial pole offsets (CPO) to estimate the period and amplitude of the FCN. It turned out that the obtained spectrum contains broad double peaks in the vicinity of the expected FCN signal (e.g.~\citet{Malkin07}) or an apparently varying period between $-425$ to $-450$ days~\citep{Schmidt05}. Figure~\ref{Fig_dXdY_fft} shows the spectrum of the CPO with a double peak around $-410$ and $-470$ days as obtained by fast Fourier transformation of our VLBI estimates from 1984.0 to 2011.0.\\
Earth rotation theory, as nowadays widely accepted, predicts one strong oscillation with a stable period. The apparent change of the period, which is seen in the spectral analysis of CPO, is attributed to a variable phase and amplitude of the rotation. The non-rigid Earth nutation model of \citet{Mathews02} is the basis for the current IAU 2000A nutation model. It predicts an FCN period between $-429.93$ and $-430.48$ solar days. The time stability of the FCN period was first examined by~\citet{Roosbeek99}, who found a period between $-431$ and $-434$ sidereal days from analysing several sub-intervals of the VLBI time series. They used the transfer function by \citet{Wahr79}, which expresses the ratio between rigid and non-rigid amplitudes of nutation terms at their  frequencies and which accounts for a resonance effect of the FCN at forced nutations. This proposal of~\citet{Roosbeek99} for an indirect estimation of the FCN period was extended by~\citet{Vondrak05} by applying the transfer function given in \citet{Mathews02} to the CPO obtained by a combined VLBI/global positioning system (GPS) solution, yielding a stable value of $-430.55~\pm~0.11$ solar days ($-431.73~\pm~0.11$ sidereal days). \citet{Lambert07a} extended the work of \citet{Vondrak05} by investigating the CPO time series from 1984.0 to 2006.0 provided by several VLBI analysis centres. They concluded that the resonant period stays stable within half a day with an average value of $-429.75~\pm~0.42$ solar days ($-430.93~\pm~0.42$ sidereal days).\\
All these studies estimated the FCN period "a posteriori", i.e. by analysing the CPO time series. In this work we use a common adjustment of the VLBI measurements for the estimation of the FCN period within a so-called global solution, where partial derivatives of the group delay $\tau$, i.e. of the primary geodetic observable of the VLBI technique, with respect to the FCN period are set up.

\section{FCN in nutation motion}
\label{sec_FCNnut}
The FCN components $X_{FCN}$ and $Y_{FCN}$ in a nutation model can be described by a time-varying sinusoidal representation:
\begin{equation}\label{XYfcn}
\begin{array}{cc}
X_{FCN} = A_{C} \cos (\sigma_{FCN} t) - A_{S} \sin (\sigma_{FCN} t),\\
Y_{FCN} = A_{S} \cos (\sigma_{FCN} t) + A_{C} \sin (\sigma_{FCN} t),
\end{array}
\end{equation}
where $A_{C}$ and $A_{S}$ are the amplitudes of the cosine and sine term, $t$ is the time given since J2000.0, and $\sigma_{FCN}$ is the frequency of FCN in the CRF.\\
In order to obtain the partial derivatives of the VLBI observable with respect to the FCN period and amplitude, the equations (\ref{XYfcn}) for FCN offsets are included into the description of the celestial motion of the CIP. The FCN offsets from equation~(\ref{XYfcn}) are simply added to the celestial pole coordinates $X_{(IAU)}$ and $Y_{(IAU)}$ following the IAU 2006/2000A precession-nutation model:

\begin{equation}\label{XYwithfcn}
\begin{array}{cc}
X=X_{FCN} + X_{(IAU)}, \\
Y=Y_{FCN} + Y_{(IAU)}.
\end{array}
\end{equation}

This addition is practically equivalent to a multiplication of the transformation matrix $Q_{(IAU)}$ \citep{iers10}:

\begin{equation}\label{Qwithfcn}
    Q(t)= dQ(t) \cdot Q(t)_{(IAU)} =
    \left[
\begin{array}{ccc}
1 & 0   & X_{FCN}\\
0   & 1 & Y_{FCN}\\
-X_{FCN}     & -Y_{FCN}     & 1
\end{array}
\right]
 \cdot Q(t)_{(IAU)}.
\end{equation}

For the combined estimation of the FCN period $P_{FCN}$ with the solid Earth tidal displacement, we express the FCN frequency in the CRF with the frequency of NDFW in the TRF $\sigma_{NDFW}$. The transformation is done by a basic relationship between frequencies in the terrestrial and celestial reference systems:

\begin{equation}\label{fcnper2}
P_{FCN} = \frac{2\pi}{\sigma_{FCN}} = \frac{1}{1-\sigma_{NDFW}} \cdot \frac{1}{sd}.
\end{equation}

It follows that $\sigma_{FCN}=2\pi \cdot sd (1-\sigma_{NDFW})$ with $sd = 1.002737909$ giving the number of sidereal days per one solar day. The partial derivatives of $dQ$ with respect to the NDFW frequency $\sigma_{NDFW}$ then read
\begin{equation}\label{pd_Qwithfcn}
   \frac{\partial dQ(t)}{\partial \sigma_{NDFW}}=
    \left[
\begin{array}{ccc}
0 & 0   & -2\pi\cdot sd \cdot t \cdot \Upsilon_{x} \\
0   & 0 & -2\pi\cdot sd \cdot t \cdot \Upsilon_{y}\\
 2\pi\cdot sd \cdot t \cdot \Upsilon_{x}     & 2\pi \cdot sd\cdot t \cdot \Upsilon_{y}    & 0
\end{array}
\right],
\end{equation}

where $\Upsilon_{x}$ and $\Upsilon_{y}$ denote

\begin{equation}\label{upsil}
\begin{array}{cc}
\Upsilon_{x} = -A_{C} \sin( \sigma_{FCN} t) -A_{S} \cos( \sigma_{FCN} t),\\
\Upsilon_{y} = -A_{S} \sin( \sigma_{FCN} t) +A_{C} \cos(\sigma_{FCN} t) .
\end{array}
\end{equation}

The partial derivatives of $dQ$ with respect to the amplitude of the cosine term $A_{C}$ are easily created as

\begin{equation}\label{pd_QAc}
   \frac{\partial dQ(t)}{\partial A_{C}}=
    \left[
\begin{array}{ccc}
0 & 0   &    \cos(\sigma_{FCN} t) \\
0   & 0 &    \sin(\sigma_{FCN} t)\\
-\cos(\sigma_{FCN} t)    &  -\sin(\sigma_{FCN} t)   & 0
\end{array}
\right],
\end{equation}

and the partial derivatives of $dQ$ with respect to the amplitude of the sine term $A_{S}$ read

\begin{equation}\label{pd_QAs}
   \frac{\partial dQ(t)}{\partial A_{S}}=
    \left[
\begin{array}{ccc}
0 & 0   &    -\sin(\sigma_{FCN} t) \\
0   & 0 &     \cos(\sigma_{FCN} t)\\
\sin(\sigma_{FCN} t)    &  -\cos(\sigma_{FCN} t)   & 0
\end{array}
\right] .
\end{equation}

The incorporation of the partial derivatives of $dQ$ into the partial derivative of the whole basic VLBI model follows as

\begin{equation}\label{twrtndfw_nut}
\frac{\partial \tau}{\partial \sigma_{NDFW}} = k(t) \cdot \frac{\partial dQ(t)}{\partial \sigma_{NDFW}} \cdot Q(t)_{(IAU)}  \cdot  R(t) \cdot W(t) \cdot b(t),
\end{equation}

where ${k}$ is the source unit vector defined in the barycentric celestial reference system, ${Q}$, ${R}$ and ${W}$ are the transformation matrices between the CRF and TRF due to nutation, Earth rotation angle, and polar motion respectively, and $b$ is the baseline vector between two VLBI stations expressed in the terrestrial reference system. In the same way one gets the partial derivative of the VLBI model with respect to the amplitude of the cosine and sine term.

\begin{table*}
\caption{Period of the FCN estimated in solutions S1 and S2, together with constant corrections to the a priori amplitudes of the FCN from~\citet{Lambert07} and to the annual and semi-annual nutation terms given in the IAU 2000A model.}
\label{table_FCN_periodCIP}
\centering
\begin{tabular}{c c c c c c c c  }
\hline\hline
Solution & $P$           & $A_{C}$  & $A_{S}$ & $A_{C}$  & $A_{S}$ & $A_{C}$  & $A_{S}$ \\
         & [sid. days]   & [$\mu$as]& [$\mu$as]  & [$\mu$as]& [$\mu$as] & [$\mu$as]& [$\mu$as]  \\
       &  & \multicolumn{2}{c}{FCN} &  \multicolumn{2}{c}{annual term} &  \multicolumn{2}{c}{semi-annual term}\\
\hline
 S1 &  $-431.17~\pm$ 0.09 & 64.6 $\pm$ 1.0 & 34.0 $\pm$ 1.2 & - & - & - & - \\
 S2 &  $-431.18~\pm$ 0.09 & 64.1 $\pm$ 1.0 & 33.9 $\pm$ 1.2 & $ -4.6 \pm$ 1.0 & $ 14.9 \pm$ 0.9 & $ -19.3 \pm$ 0.9 & $ -8.9 \pm$ 0.9   \\
\hline
\end{tabular}
\end{table*}

\subsection{Analysis of the VLBI measurements}
\label{cpo_vlbi}
We estimated the FCN period from the motion of the CIP in the geocentric celestial reference system (GCRS) as a global parameter in a common adjustment (global solution) of 3360 24-hour sessions of the International VLBI Service for Geodesy and Astrometry (IVS) \citep{Schuh12}. These sessions fulfil two criteria: a) the network is built with at least three stations, and b) the a~posteriori sigma of unit weight obtained from a single-session adjustment does not exceed the value of~2. The whole analysis of 4.6~million observations from 1984.0 to 2011.0 was done with the Vienna VLBI Software (VieVS) \citep{Boehm12}. The theoretical time delays were modelled according to recent IERS Conventions 2010, with the exception of applying a priori corrections on station coordinates due to non-tidal atmosphere loading \citep{Petrov04}, which is a common procedure in VLBI analysis. The celestial motion of the CIP was modelled according to equation~(\ref{XYwithfcn}). The FCN offsets were taken from the model by~\citet{Lambert07}, who uses the a priori FCN period of $-431.39$~sidereal days by~\citet{Mathews02} and provides the amplitude terms $A_{C}$ and $A_{S}$ as determined empirically from the CPO in the IERS EOP05 C04 combined series. The values of $A_{C}$ and $A_{S}$ are given in yearly steps and the amplitudes during the year are obtained by linear interpolation.\\
The VieVS was extended with partial derivatives of the measured time delay with respect to the FCN period as described in equation~(\ref{twrtndfw_nut}) and to the FCN amplitude. Furthermore, partial derivatives with respect to the annual and semi-annual harmonic terms in the nutation motion were added.\\
Two solutions were run with the same a priori parameterisation. In both solutions a new TRF and a new CRF were estimated as global parameters by applying no-net-translation and no-net-rotation conditions with respect to VTRF2008~\citep{Boeckmann10} and ICRF2~\citep{Fey09} respectively. Clock parameters, zenith wet delays, tropospheric parameters, and Earth rotation parameters were session-wise reduced.
\begin{itemize}
     \item In solution S1 the FCN period together with the constant corrections to the cosine and sine amplitude terms were estimated as global parameters.
     \item Solution S2 is identical to solution S1, but additional cosine and sine amplitudes of the annual and semi-annual harmonic terms in nutation were determined.
\end{itemize}

Due to the non-linear relationship of the FCN period in the FCN offsets, several iterative solutions had to be run. In solution S1 the period of FCN in the global solution is estimated as $-431.17~\pm~0.09$~sidereal days and the amplitude corrections are $64.6~\pm~1.0~\mu$as for the cosine term and $34.0~\pm~1.2~\mu$as for the sine term. The resulting FCN period obtained from solution S2 ($-431.18~\pm$ 0.09~sidereal days) is almost identical to the estimates from solution~S1. The values of the remaining absolute amplitudes of the annual and semi-annual terms (in addition to the values included in the IAU 2000A nutation model) are $15.6~\pm~1.0~\mu$as and $21.3~\pm~1.0~\mu$as respectively. The amplitude value from the cosine and sine terms is obtained in the usual way as $A = \sqrt{(A_{C}^2 + A_{S}^2)}$. The comparison of solutions S1 and S2 shows that an additional estimation of corrections to the annual and semi-annual nutation terms does not influence the FCN period determination. The FCN period from solutions S1 and S2 with the constant corrections to the cosine and sine amplitude terms for the FCN and the annual and semi-annual nutation terms are summarised in Table~\ref{table_FCN_periodCIP}.

\section{FCN in solid Earth tides }
The FCN affects the solid Earth tides in their diurnal band, causing a strong resonance effect. The Love and Shida numbers, i.e. the proportionality parameters between the tide-generating potential and the tidal displacement, for the diurnal tidal waves in the vicinity of the NDFW period depend on frequency, see e.g.~\citet{Krasna13}. We use the resonance effect in these tidal waves to determine the FCN period directly from VLBI analysis, which was first done by~\citet{Haas96}. Love and Shida numbers in the diurnal band can be represented by a resonance formula as a function of the tidal excitation frequencies with the frequency of Chandler wobble $\sigma_{CW}$, of the NDFW $\sigma_{NDFW}$, and of the free inner core nutation (FICN) $\sigma_{FICN}$ \citep{Mathews95, iers10}:

\begin{equation}\label{resdiu}
    L_f = L_0 + \frac{L_{CW}}{\sigma_f-\sigma_{CW}} + \frac{L_{NDFW}}{\sigma_f-\sigma_{NDFW}}+ \frac{L_{FICN}}{\sigma_f-\sigma_{FICN}},
\end{equation}

where $L_f$ is a generic symbol for the frequency-dependent Love ($h$) and Shida ($l$) numbers, with $L_0$, $L_{CW}$, $L_{NDFW}$, and $L_{FICN}$ as resonance coefficients \citep{iers10}. In the terrestrial diurnal band only the periods of the NDFW and the FICN can be found. The principal resonance comes from the NDFW with a resonance strength factor ($L_{NDFW} = 0.18053 \cdot 10^{-3}$) 100~times larger than that of the FICN ($-0.18616 \cdot 10^{-5}$). The partial derivative of the station displacement in the local coordinate system with respect to the NDFW frequency follows from the frequency-dependent corrections $\delta{d_f}$ to the displacement vector, which can be written as \citep{iers10}

\begin{equation}\label{diu_ren_f}
\begin{aligned}
\delta{d_f}=-3\sqrt{\frac{5}{24\pi}}H_f \Bigg\{ & \delta{h_f}\frac{1}{2}\sin2\Phi\sin(\theta_f+\Lambda)\;\hat{r} \\
                          +\; &\delta{l_f}\sin\Phi\cos(\theta_f+\Lambda)\;\hat{e}     \\
                          +\; &\delta{l_f}\cos2\Phi\sin(\theta_f+\Lambda)\;\hat{n}   \Bigg\},
\end{aligned}
\end{equation}

where $\delta{h_f}$ and $\delta{l_f}$ are the corrections to the constant values of Love and Shida numbers $h_2$ and $l_2$, which equal to 0.6078 and 0.0847 respectively, according to \citet{iers10}; $H_f$ is the amplitude of the tidal term with frequency $f$ using the defining convention by \citet{Cartwright71}; $\Phi$ and $\Lambda$ are the geocentric latitude and longitude of the station; $\theta_f$ is the tide argument for tidal constituent with frequency $f$; $\hat{r},\hat{e},\hat{n}$ are unit vectors in radial, east, and north direction respectively.

The partial derivative of the basic VLBI model with respect to the NDFW frequency contained in the solid Earth tides, i.e. in the displacement of stations building a baseline, is in its general form given by equation~(\ref{twrtndfw_bas}):

\begin{equation}\label{twrtndfw_bas}
\begin{aligned}
\frac{\partial \tau}{\partial \sigma_{NDFW}} &= k(t) \cdot Q(t) \cdot  R(t) \cdot W(t) \cdot \frac{\partial b(t)}{\partial \sigma_{NDFW}}.
\end{aligned}
\end{equation}

For the analysis of the VLBI measurements, the same a priori modelling and parameterisation as described in section~\ref{cpo_vlbi} were applied. The FCN period was obtained together with a simultaneously estimated TRF and CRF. After four iterative runs the period stayed stable at $-431.23~\pm~2.44$ sidereal days.

\begin{table*}
\caption{Cosine and sine amplitude terms of the FCN model determined in yearly steps within global solutions of VLBI measurements.}
\label{table_FCNmodel}
\centering
\begin{tabular}{r r r | r r r | r r r }
\hline\hline
Year & $A_{C}$ [$\mu$as]& $A_{S}$ [$\mu$as] & Year & $A_{C}$ [$\mu$as]& $A_{S}$ [$\mu$as] & Year & $A_{C}$ [$\mu$as]& $A_{S}$ [$\mu$as]\\
\hline
1986.0 &   $-256.6 \pm 9.8$    &    $-162.6 \pm 9.8 $  &    1994.0 &   $-108.3 \pm 2.6$    &    $  19.7 \pm 2.6 $   &   2002.0 &   $  98.4 \pm 2.0$    &    $ -82.9 \pm 2.0 $    \\
1987.0 &   $-261.1 \pm 9.1$    &    $-104.3 \pm 9.1 $  &    1995.0 &   $-105.2 \pm 2.1$    &    $  17.7 \pm 2.2 $   &   2003.0 &   $ 104.5 \pm 1.9$    &    $ -71.0 \pm 1.9 $    \\
1988.0 &   $-216.3 \pm 9.1$    &    $ -84.9 \pm 9.1 $  &    1996.0 &   $ -99.4 \pm 2.2$    &    $  18.3 \pm 2.2 $   &   2004.0 &   $ 109.0 \pm 1.8$    &    $ -56.2 \pm 1.7 $    \\
1989.0 &   $-180.5 \pm 7.5$    &    $ -45.6 \pm 7.5 $  &    1997.0 &   $ -89.9 \pm 2.3$    &    $  16.9 \pm 2.3 $   &   2005.0 &   $ 111.9 \pm 2.0$    &    $ -23.0 \pm 2.0 $    \\
1990.0 &   $-166.0 \pm 6.1$    &    $  -6.3 \pm 6.1 $  &    1998.0 &   $ -76.0 \pm 2.4$    &    $   2.8 \pm 2.4 $   &   2006.0 &   $ 121.1 \pm 1.8$    &    $  25.4 \pm 1.8 $    \\
1991.0 &   $-145.3 \pm 5.0$    &    $  19.8 \pm 5.0 $  &    1999.0 &   $ -39.8 \pm 2.7$    &    $ -32.2 \pm 2.8 $   &   2007.0 &   $ 150.3 \pm 1.7$    &    $  75.5 \pm 1.7 $    \\
1992.0 &   $-146.3 \pm 4.0$    &    $  26.7 \pm 3.9 $  &    2000.0 &   $   8.3 \pm 2.6$    &    $ -82.3 \pm 2.6 $   &   2008.0 &   $ 162.1 \pm 1.8$    &    $ 134.0 \pm 1.8 $    \\
1993.0 &   $-128.7 \pm 3.0$    &    $  23.5 \pm 3.0 $  &    2001.0 &   $  57.7 \pm 2.3$    &    $-102.4 \pm 2.3 $   &   2009.0 &   $ 145.8 \pm 2.2$    &    $ 156.3 \pm 2.2 $    \\
\hline
\end{tabular}
\end{table*}

\section{Simultaneous estimation of the FCN period from solid Earth tides and nutation}
\label{FCNperiod_both}
In previous sections the presence and effects of the FCN in the solid Earth tides and in the nutation of the CIP were treated separately. In this part we introduce a rigorous determination of the FCN period, where the partial derivative of the observation equation contains changes in both parameters (nutation matrix and baseline vector), which are influenced by the presence of the FCN:

\begin{equation}\label{fcn_joined}
\begin{aligned}
\frac{\partial \tau}{\partial \sigma_{NDFW}}
&= k(t) \cdot \frac{\partial dQ(t)}{\partial \sigma_{NDFW}} \cdot Q(t)_{(IAU)}  \cdot  R(t) \cdot W(t) \cdot b(t) \\
&+ k(t) \cdot Q(t) \cdot R(t) \cdot W(t) \cdot \frac{\partial b(t)}{\partial \sigma_{NDFW}}.
\end{aligned}
\end{equation}

The treatment of the FCN in the CIP motion agrees with solution S1 in section \ref{sec_FCNnut}, i.e. a priori values for the FCN period and amplitudes are taken from the model of~\citet{Lambert07}. Constant offsets to the sine and cosine amplitudes over the 27~years of VLBI data are  estimated in the global adjustment. Other globally estimated parameters are the TRF and CRF. The estimate of the FCN period after four iterations is $-431.18~\pm~0.10$ sidereal days, which is very close to the result from the "nutation only" solution. We assume that the highly precise estimation of the FCN period from nutation motion is achieved by the direct observation of the FCN in the rotation motion of the Earth axis. The less precise estimate of the FCN period obtained from the station displacement may reflect the indirect resonance effect on the solid Earth tidal motion.

\section{Empirical FCN model with globally estimated varying amplitude}
\begin{figure}
\centering
\includegraphics[width=\hsize]{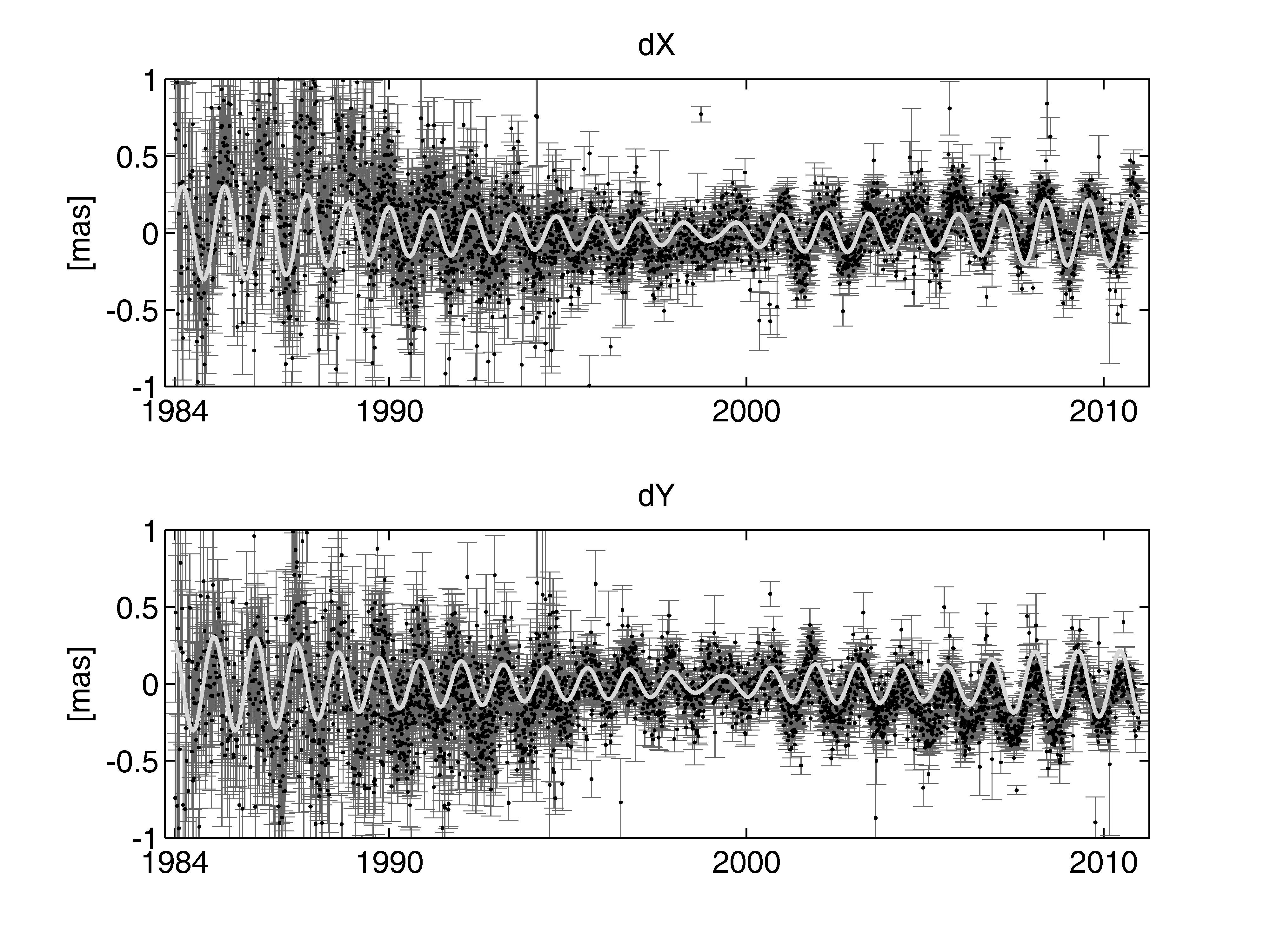}
  \caption{CPO with respect to the IAU 2006/2000A precession-nutation model (grey) together with the FCN model (light grey) estimated in this work. Before 1986.0 and after 2009.0, the model is extrapolated.}
     \label{Fig_dXdYmodel}
\end{figure}

In section~\ref{sec_FCNnut} the FCN model created by~\citet{Lambert07} was introduced where the time-varying amplitudes (cosine and sine terms) were fitted through the CPO in IERS EOP 05 C04 combined series with a sliding window over two years and displaced by one year. Following this idea of a varying amplitude and phase estimated in a one-year step, we determined the amplitudes $A_{C}$ and $A_{S}$ in several global solutions. The data input for each run are VLBI measurements carried out over four years, starting in 1984.0. Estimated parameters are constant cosine and sine amplitude terms corresponding to the FCN period of $-431.18$ sidereal days, as estimated in the joint adjustment described in section~\ref{FCNperiod_both}. The partial derivatives are given by equations~(\ref{pd_QAc}) and~(\ref{pd_QAs}) and the estimates of the $A_{C}$ and $A_{S}$ refer to the middle of the analysed data spans. Other estimated parameters are the session-wise reduced clock parameters, zenith wet delays, tropospheric gradients, and Earth rotation parameters. The TRF and CRF are fixed to the reference frames estimated in section~\ref{FCNperiod_both} to avoid a different datum definition dependent on included stations and radio sources in the respective groups of four years' measurement data. In the first run data from 1984.0 to 1988.0 were involved and thus the estimates are valid for 1986.0. The second global solution includes data from 1985.0 till 1989.0, and it continues to the year 2011.0. The estimated values are shown in Table~\ref{table_FCNmodel} and the resulting FCN model is plotted in Figure~\ref{Fig_dXdYmodel}. We found a very good agreement between our solution and the one from~\citet{Lambert07} with differences in the cosine and sine amplitudes smaller than several microarcseconds.

\section{Conclusions}
The FCN period is estimated within a global VLBI solution from solid Earth tidal displacement as $-431.23~\pm~2.44$ sidereal days and from the motion of the CIP as $-431.17~\pm~0.09$ sidereal days, together with constant sine and cosine amplitude terms. The final value for the FCN period is derived from the solid Earth tidal displacement and from the motion of the CIP in a joint solution. Its estimated value of $-431.18~\pm~0.10$ sidereal days differs slightly from the conventional value $-431.39$ sidereal days given in~\citet{iers10}. Furthermore, we present new values of an empirical FCN model. The period is fixed to the value determined in our joint solution, and the cosine and sine amplitudes are estimated from several global solutions in yearly steps directly from VLBI measurements.

\begin{acknowledgements}
      The authors acknowledge the International VLBI Service for Geodesy and Astrometry (IVS) \citep{Schuh12} and all its components for providing VLBI data. H. Kr{\'a}sn{\'a} works within FWF-Project P23143-N21 (Integrated VLBI). The authors would like to thank the referee S. Lambert for his valuable comments.
\end{acknowledgements}

\bibliographystyle{aa}

\bibliography{references_krasna_AA_2013_21585}

\end{document}